\documentclass[aps,twocolumn,tightenlines]{revtex4}
\usepackage{graphicx}
\def\wmap{{\sl WMAP\ }}
\begin{document}
\bibliographystyle{apsrev}

\title{Constraining the Topology of the Universe}
\author{Neil J. Cornish}
\affiliation{Department of Physics, Montana State University, Bozeman, MT 59717}
\author{David N. Spergel}
\affiliation{Department of Astrophysical Sciences, Princeton University, Princeton, NJ 08544}
\author{Glenn D. Starkman}
\affiliation{Center for Education and Research in Cosmology and Astrophysics,
Department of Physics, Case Western Reserve University, Cleveland, OH 44106--7079 \\
and CERN Theory Division, CH--1211 Geneva 23, Switzerland}
\author{Eiichiro Komatsu}
\affiliation{Department of Astrophysical Sciences, Princeton University, Princeton, NJ 08544,   \\
Department of Physics, Princeton University, Princeton NJ 08544 \\
and Department of Astronomy, The University of Texas at Austin, Austin, TX 78712}
\begin{abstract}
The first year data from the Wilkinson Microwave Anisotropy Probe are
used to place stringent constraints on the topology of the Universe.
We search for pairs of circles on the sky with similar temperature patterns
along each circle. We restrict the search to back-to-back circle pairs,
and to nearly back-to-back circle pairs, as this covers the majority of the
topologies that one might hope to detect in a nearly flat universe. We do
not find any matched circles with radius greater than 25$^\circ$.  For a
wide class of models, the non-detection rules out the possibility that we
live in a universe with topology scale smaller than 24 Gpc.
\end{abstract}

\maketitle

Recently, the Wilkinson Microwave Anisotropy Probe ({\sl WMAP})  has 
produced a high resolution, 
low noise map of the temperature fluctuations in the Cosmic Microwave
Background (CMB) radiation~\cite{bennett1}.
Our goal is to place constraints on the topology of
the Universe by searching for matched circles in this map. 
Here we report the results
of a directed search for the most probable topologies. 
The results of the full search,
and additional details about our methodology, will 
be the subject of a subsequent publication.

The question we seek to address can be plainly stated: 
Is the Universe finite or infinite? Or,
more precisely, what is the shape of space? 
Our technique for studying the shape, or topology,
of the Universe is based on 
the simple observation that if we live in a Universe that is
finite, light from a distant object will be 
able to reach us along more than one path.
The one caveat is that the light must have sufficient time to 
reach us from multiple directions,
or put another way, that the Universe is sufficiently small. 
The idea that space might be
curled up in some complicated fashion has a long history.
In 1900 Schwarzschild~\cite{ks} considered
the possibility that space may have non-trivial topology, and used the multiple image idea to
place lower bounds on the size of the Universe. Recent
progress is summarized in Ref.~\cite{janna}.    

The results from the \wmap
experiment~\cite{bennett1} have deepened interest in the 
possibility of a finite universe. Several reported large scale
anomalies are all potential signatures of a finite universe:
the lack of large angle fluctuations~\cite{spergel}, reported
non-Gaussian features in the maps~\cite{oliveira,eriksen}, and features
in the power spectrum~\cite{luminet}.

The local geometry of space constrains, but does not dictate the topology of space.
A host of astronomical observations support the idea that space is 
locally homogeneous and isotropic, 
so we may restrict our attention to the three dimensional spaces of constant
curvature: Euclidean space $E^3$; Hyperbolic space $H^3$; 
and Spherical space $S^3$.
A useful way to view non-trivial topologies with 
these local geometries is to imagine
the space being tiled by identical copies of a fundamental cell. 
For example, Euclidean space can be tiled by cubes, 
resulting in a three-torus topology. Assuming that light
has sufficient time to cross the fundamental cell, 
an observer would see multiple copies
of a single astronomical object. 
To have the best chance of seeing ``around the Universe''
we should look for multiple images of the most 
distant reaches of the Universe. The
last scattering surface, or decoupling surface, marks the 
edge of the visible Universe.
Thus, looking for multiple imaging of the last scattering surface
is a powerful way to look for non-trivial topology.

But how can we tell that the essentially random 
pattern of hot and cold spots on the
last scattering surface have been multiply imaged? 
First of all, the microwave photons
seen by an observer have all been traveling at the same speed, 
for the same amount of time, 
so the the surface of last scatter is 2-sphere centered on the observer. 
Each copy of the observer will come with a copy of the
surface of last scatter, and if the copies are separated by a distance
less than the diameter of the surface of last scatter, then
the copies of the surface of last scatter will intersect. 
Since the intersection of two 2-spheres defines a circle,
the surfaces of last scatter will intersect along circles. 
These circles are visible by both copies of the observer, but from opposite sides.
Of course the two copies are really one observer, so
if space is sufficiently small, 
the cosmic microwave background radiation from the surface of last
scatter will 
have patterns of hot and cold spots that match around circles~\cite{us}.
The key assumption in this analysis
is that the CMB fluctuations come primarily from
the surface-of-last scatter and are due to density and potential
terms at the surface of last scatter.

Implementing our matched circle test is  straightforward
but computationally intensive.
The general search must explore a six dimensional parameter space. 
The parameters are the location of the first circle center, $(\theta_1,\phi_1)$, 
the location of the second circle center, $(\theta_2, \phi_2)$, 
the angular radius of the circle $\alpha$,
and the relative phase of the two circles $\beta$. 
We use the Healpix scheme to define
our search grid on the sky. 
A resolution $r$ Healpix grid divides the sky into
$N=12 N_{side}^2$ equal area pixels, where $N_{side}=2^r$.
Similar angular resolutions were used to step
through $\alpha$ and $\beta$. 
Thus, the total number of circles being compared scales as $\sim N^3$,
and each comparison takes $\sim N^{1/2}$ operations. 
The simplest implementation of the search at the resolution of
the \wmap data ($r=9$) would take greater than $10^{20}$ operations.
This is not computationally feasible. However, with the algorithms
outlined below, we can carry out a complete search for most
likely topologies using the \wmap data.

The \wmap data suggest that the Universe is very nearly spatially flat, with
a density parameter $\Omega_0 = 1.02\pm 0.02$\cite{spergel}. Our universe
is either
Euclidean, or 
its radius of curvature is large compared to radius of the surface of
last scatter. 
For topology to be observable using our matched circle technique we require
that the distance to our nearest copy 
is less than the 
diameter of the surface of
last scatter, which in turn implies that, near 
our location and in at least one direction,
the fundamental cell is small compared to the radius of curvature. 
Given the observational constraint on the curvature radius,
it is highly unlikely that there are any 
hyperbolic topologies small enough to be detectable~\cite{jeff1},
and there are strong constraints on the 
types of spherical topologies that might be
detected~\cite{jeff2}. Naturally, the near flatness of the Universe does 
not place any
restrictions on the observability of the Euclidean topologies. 
Remarkably, the largest matching circles in most of the topologies we might hope 
to detect will be back-to-back on the sky or nearly so. This immediately 
reduces the search space from six to
four dimensions. 
This result is exact and easy to prove for nine of the ten Euclidean topologies.
The one exception is the Hantzsche-Wendt space, which has its largest circles with
centers separated by 90$^\circ$.
The result is less obvious in spherical space~\cite{jeff2}
but is nonetheless exact for a large class of such spaces (single action manifolds).  
All others (double or linked action manifolds) 
predict slightly less than 180$^\circ$  separations between the circle centers
for generic locations of the observer, but larger offsets for special locations.
This generically slight offset motivates the final
search described in this letter.  
The possibility of a large offset requires the full search which will be described in a future 
publication.

How do we compare the circles?  Around each pixel $i$, 
we draw a circle of radius  $\alpha$ and linearly interpolate values at $n = 2^{r+1}$ points along
the circle.  We then Fourier transform each circle:
$T_{i}(\phi) = \sum_m T_{im} \exp{i m \phi}$ and compare circle
pairs, with equal weight for all angular scales:
\begin{equation}
S_{ij}(\alpha,\beta) = \frac{ 2 \sum_m m T_{im}(\alpha) T^*_{jm}(\alpha) e^{-i m \beta}}
{ \sum_m m \left( \vert  T_{im}(\alpha) \vert^2 + \vert T_{jm}(\alpha) \vert^2 \right)} \, .
\end{equation}
The $i,j$ refer to the location of the circle centers. With
this definition $S_{ij} = 1$ for a perfect match.  For two
random circles $<S_{ij}> = 0$.  We estimate $<S_{ij}^2>$ below.
We can speed the calculation of the $S$ statistic by rewriting it as
\begin{equation}
S_{ij}(\alpha,\beta) = \sum_m s_m e^{-i m \beta} \, ,
\end{equation}
where
\begin{equation}
s_m = \frac{2 m T^*_{im} T_{jm}}
{\sum_n n\left( \vert  T_{in} \vert^2 + \vert T_{jn} \vert^2 \right)} \, .
\end{equation}
By performing an inverse fast Fourier transform of the $s_m$, we get $S_{ij}(\alpha,\beta)$
at a cost of $\sim N^{1/2} \log N$ operations. This reduces the cost of the back-to-back search
from $\sim N^{5/2}$ to $\sim N^2 \log N$. We further speed the calculation
by using a hierarchical approach: we first search at $r = 7$ and identify
the 5000 best circles at a fixed value of $\alpha$.  We then refine
the search around these best pixel centers and radii and complete
the search at $r=9$.  We store the best 5000 matches at each $\alpha$.
$S_{max}(\alpha)$ is the best match value found at each radius.

To test the search algorithm, we generate a simulated CMB sky for a finite
universe. 
When only the Sachs-Wolfe effect \cite{sachs} was included, the algorithm
found nearly perfect circles:
$S_{max} = 0.99$ with pixelization effects accounting for the residual errors.
Real world effects, however, significantly degrade the quality of the circles.  The dominant
source of noise are velocities at the surface of last scatter which 
dilutes the quality of the expected match.  Following the approach
outlined in the appendix of~\cite{komatsu}, we generate CMB skies with
initial Gaussian random amplitudes for each $k$ mode and then compute
the transfer function for each value of $k$.  This approach implicitly
includes all of the key physics: finite surface of last scatter, reionization,
the ISW effect and the Doppler term.  
Since topology
has little effect on the power spectrum for multipoles $l > 20$, we use the best
fit parameters based on the analysis of the \wmap temperature and
polarization data~\cite{spergel}.
The ratio of the amplitude of the potential and density terms (which
contribute to the signal for the circle statistic) to the velocity
and ISW terms (which contribute to the noise for the circle statistic)
depends weakly on the basic cosmological parameters: within
the range consistent with the \wmap data, the variation is
small.  For the simulated sky,
we assume that the primordial potential fluctuations had the 
usual $1/k^{3}$ power spectrum.  
The simulation included detector noise~\cite{hinshaw} and the effects
of the \wmap beams
\cite{page}.  The simulation did not include gravitational lensing
of the CMB as the lensing deflection angle in the standard
cosmology is small~\cite{seljak}, $\sim 6'$,
compared to the smoothing scale used in the analysis, $40'$.
When the search was performed on these realistic skies,
the quality of circle matches were degraded.  Figure 1 shows the results
of the search on the simulated sky: the peaks in the plots correspond to
radii at which a matched circle was detected. The largest circles had
$S_{max}=0.75$, and the best match value decreased for smaller circles. 
The code was able to find all 107
predicted circle pairs with radii greater than 25$^\circ$: 
the poorest detection had
an $S_{max}$ value of 0.45.  While we only simulated the three torus,
the degradation of the circle match as a function of radius will
be nearly identical for other topologies as the contribution of
the Doppler term (and detector noise) is determined by local physics
at angular scales well below the circle radius.

\begin{figure}[t]
\vspace{68mm}
\includegraphics{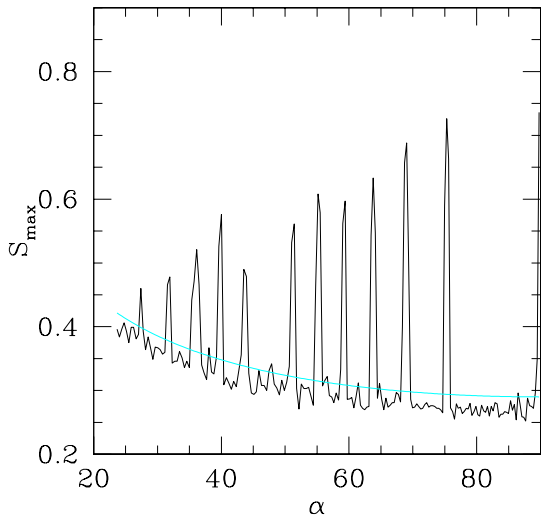}
\caption{The maximum value of the circle statistic as a function
of radius, $\alpha$, for a simulated finite universe model.  The peaks in
the plots correspond to positions of matched circles. The cyan line
corresponds to the detection threshold discussed below. $\alpha$
is measured in degrees in all figures.
$S_{max}(\alpha)$ is the best match value found at each radius.}
\end{figure}

For the circle search, we generated a low foreground CMB map from the \wmap data.
Outside the \wmap Kp2 cut, we used a noise-weighted combination of the \wmap
Q, V and W band maps. Using the template approach of~\cite{bennett2}, these
maps were corrected for dust, foreground and synchrotron emission.  
This map was smoothed to the resolution of the Q band map.
Inside the Kp2 cut, we used the internal linear combination map~\cite{bennett2}.

What is the expected level of the false positive signal?  
If we approximate the CMB sky as a Gaussian random field with
coherence angle $\theta_c$, then there are roughly $N_{circ}= 2\pi/\theta_c^2$
independent circles on the sky of radius $\alpha$.  Along
each circle, there are $2\pi \sin(\alpha)/\theta_c$ independent
patches and also $2\pi \sin(\alpha)/\theta_c$ independent
orientations. Thus, the back-to-back search considered $
N_{search} = 4 \pi^2 \sin(\alpha)/\theta_c^3$
possible circle pairs.  If we treat the patches as independent, then 
$<S^2>^{1/2} = 1/\sqrt{2 \pi \sin(\alpha)/\theta_c}$.  
The number of circles expected above a given threshold is,
$N(S > S_0) = N_{search}/2 {\rm erfc}(S_0/\sqrt{2 <S^2>})$. 
Thus, inverting the complementary error function yields the maximum value for any false positive circle at each radius,
\begin{equation}
S_{max}^{f.p.}(\alpha) \simeq <S^2>^{1/2} \sqrt{2\ln \left(\frac{N_{search}}
		{2 \sqrt{\pi \ln(N_{search})}}\right)}\label{eq:Sfp}
\end{equation}
For both the simulations and the data, 
$\theta_c \sim 0.7^\circ$. Substituting the
expression for $N_{search}$ into equation (\ref{eq:Sfp}) yields:
$S_{max}^{f.p.} \sim 0.24\sqrt{1+\ln(\sin \alpha)/18}/\sqrt{\sin\alpha}$,   This estimate
is sensitive to $\theta_c$ and excludes correlations between circles
\begin{figure}[t]
\vspace{68mm}
\includegraphics{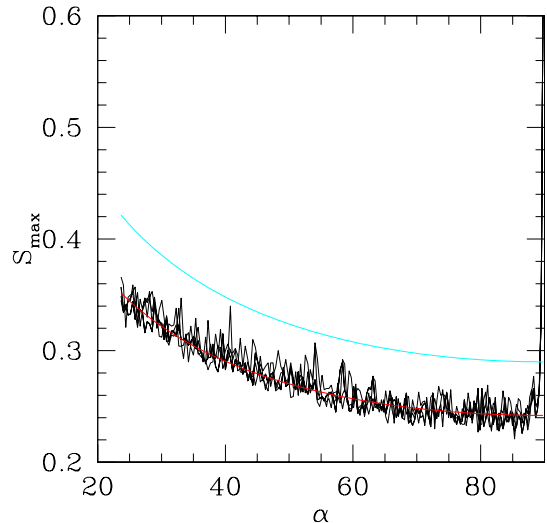}
\caption{The maximum value of the circle statistic as a function
of radius for back-to-back circle pairs on scrambled CMB skies.
The results are shown for seven different realizations.  The red
line is a fit for the false positive rate.
The cyan line shows the false detection threshold set so that fewer than 1 in 
100 simulations
would yield a false event.}
\end{figure}

We use ``scrambled" versions of
the \wmap cleaned sky map to obtain a more accurate estimate
of the false positive rate.  We generate ``scrambled" versions
of the true sky by taking the spherical harmonic transform of
the map and then randomly exchanging $a_{lm}$ values at fixed $l$.
This scrambling generates new maps with the same two point function
but different phase correlations.  The $S_{max}$ value for the best fit back-to-back
circle found at each radius is plotted as  a function of $\alpha$ in figure 2.
The false positive signal is well fit by a slightly modified form of the analytical estimates.
Based on the analytical estimate, we set a detection threshold so that fewer than 1 in
100 random skies should generate a false match. 
This threshold is shown in figures (1), (2) and (3).  

\begin{figure}[t]
\vspace{68mm}
\includegraphics{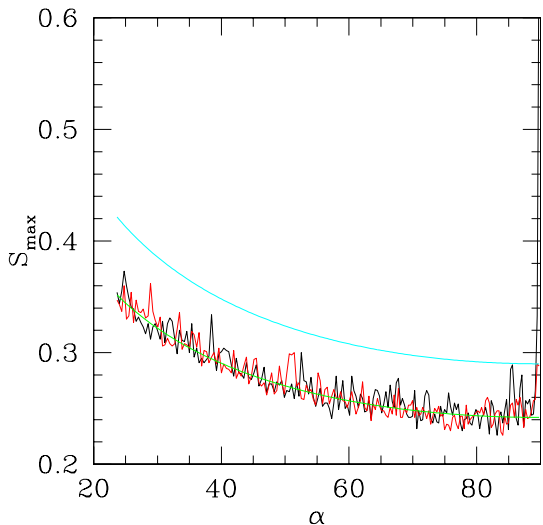}
\caption{The maximum value of the circle statistic as a function
of radius for the \wmap data for back-to-back circles.
The solid line in the figure is 
for a orientable topology.  The red line is  for non-oriented topologies.
The green line is the expected false detection level. The cyan line is the detection threshold.
The spike at $90^\circ$ is due to a trivial match between
a circle of radius $90^\circ$ centered around a point and a copy of the same
circle centered around its antipodal point.}
\end{figure}

Figure 3 shows the result of the back-to-back search on the \wmap data.
The search checked circle pairs that were both matched in phase
(points along both circle labeled in a clockwise direction) and
flipped in phase.  These two cases correspond to searches for non-orientable
and orientable topologies.
We did not detect any statistically significant circle matches
in either search.  The 
simulations show that the minimum signal expected for a circle of radius
greater than 25$^\circ$ is $S_{max}=0.45$, which is above the false detection
threshold.  Since we did not detect this signal, we can
exclude any universe that predicts circles larger than this critical size.
In a broad class of topologies,
this constrains the minimum translation distance, $d$:
\begin{equation}
d = 2 R_c {\rm atan}\left[\tan(R_{lss}/R_c) \cos(\alpha_{min})\right]
\end{equation}
where $R_c$ is the radius of curvature, and
$R_{lss}$ is the distance to the last scattering surface. Note
that one recovers the Euclidean formula for $R_c \rightarrow \infty$ and the hyperbolic
formula for $R_c \rightarrow i R_c$. For a flat or nearly flat universe, models with
$d < 2R_{lss} \cos (\alpha) \simeq 24$  Gpc are excluded.
Note that the search excludes the Poincar\'e Dodecohedron suggested
in \cite{luminet} as this model predicts back-to-back
circles of radius $35^\circ$.

\begin{figure}[t]
\vspace{68mm}
\includegraphics{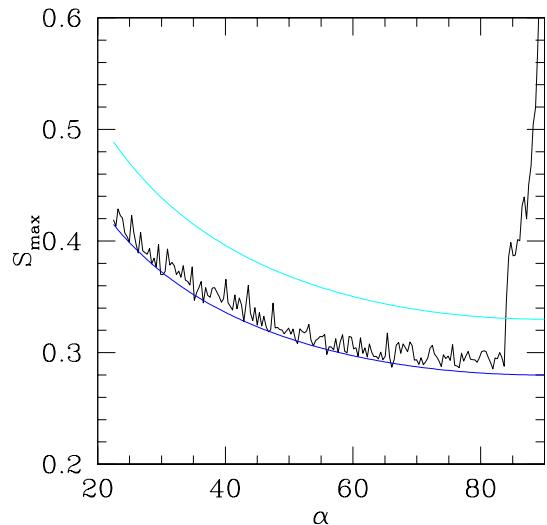}
\caption{The maximum value of the circle statistic as a function
of radius for the \wmap data for circles whose centers are separated
by greater than 170$^\circ$. The solid line in the figure is 
for a orientable topology. 
The blue line is the expected value of $S_{max}$ for a Gaussian sky (based
on  the analytic estimate and extrapolation 
from the scrambled sky simulations). The cyan line is the detection threshold.}
\end{figure}

By considering a broader set of circles, we can extend the search
to other possible topologies.  Figure 4 shows the result of
a search for nearly back-to-back circles: circles whose centers are
separated by more than 170$^\circ$.  Even this restricted search requires
searching far more circles than the back-to-back search: the search takes
4 processor months on a SGI Origin computer. Because the number of independent
circles in this search is roughly 200 times larger than in the back-to-back search,
the level of the false positive signal is higher.  The increase
in $N_{search}$ raises the false detection threshold from 
5.41 to 6.29  $<S^2>^{1/2}$:
$S^{fp}_{max} = 0.28 \sqrt{1+\ln(\sin \alpha)/6}/\sqrt{\sin\alpha}$.  Because of this
higher signal, the minimum detectable angle is 28$^\circ$ for this topology.
The signal at $\alpha \sim 85 - 90^\circ$ is due to nearly overlapping and osculating circles.
This feature is also seen in scrambled skies.
 
Has this search ruled out the possibility that we live in a finite universe?
No, it has only ruled out a broad class of finite universe models smaller
than a characteristic size.  By extending the search to all possible
orientations, we will be able to exclude the possibility that we live
in a universe smaller than 24 Gpc in diameter. More directed searches could extend
the result for specific manifolds somewhat beyond 24 Gpc.
With lower noise and higher resolution CMB maps (from \wmap's extended mission and from Planck),
we will be able to search for smaller circles and extend the limit to
$\sim$ 28 Gpc.  If the universe is larger than this, the circle
statistic will not be able to constrain it shape.


We would like to thank the \wmap team for producing wonderful
sky maps and providing insightful comments on the draft. We thank the
LAMBDA data center provided the data for the analysis. We acknowledge
the use of the following helpful software packages: HEALPIX, CMBFAST 
and Waterloo Maple.  GDS is supported by DOE Grant DEFG0295ER40898 at
CWRU.

\end{document}